\documentclass[12pt]{article}
\usepackage{amssymb} 
\def\href#1#2{#2}
\def\IP{\relax{\rm I\kern-.18em P}}

\def\be{ \begin{equation}}          \def\ee{ \end{equation}}
\def\ba{ \begin{eqnarray}}          \def\ea{ \end{eqnarray}}

 \def\Z{\mathbb{Z}} \def\R{\mathbb{R}}

\def\cedille#1{\setbox0=\hbox{#1}\ifdim\ht0=1ex \accent'30 #1%
 \else{\ooalign{\hidewidth\char'30\hidewidth\crcr\unbox0}}\fi}
\def\gaw{Gaw\cedille edzki}

\def\Ad{\mbox{\rm Ad}} \def\ad{\mbox{\rm ad}}

\def\a{\alpha }

\def\ew{\hspace*{-1mm}}   \def\ppe{\hspace*{-2.5mm}}

\newcommand{\note}[1]{\raisebox{1ex}{{\footnotesize \sf #1}}}
\newcommand{\rnote}[1]{\raisebox{1ex}{{\hspace*{-3mm} \scriptsize\sf#1}}
                       \hspace*{-4mm}}
\newcommand{\Fus}[6]{F_{{\scriptstyle #1}{\scriptstyle #2}}
  \hspace*{.3mm}\displaystyle{[} \ew \begin{array}{ll} {\scriptstyle #3 }
  \ppe & {\scriptstyle #4} \ppe \\[-2mm] {\scriptstyle #5}\ppe &
  {\scriptstyle #6}\ew \end{array}\displaystyle{]}}
\newcommand{\CG}[6]{\displaystyle{[} \,\ew \begin{array}{lll} 
  {\scriptstyle #1} \ppe
  & {\scriptstyle #2} \ppe & {\scriptstyle #3} \ew \\[-2mm] {\scriptstyle
  #4} \ppe & {\scriptstyle #5}\ppe & {\scriptstyle #6} \ew\end{array}
  \displaystyle{]}}
\newcommand{\SJS}[6]{ \displaystyle{\{ } \ew \begin{array}{lll} 
  {\scriptstyle #1} \ppe  &
  {\scriptstyle #2} \ppe & {\scriptstyle #3}
  \ppe \\[-2mm]{\scriptstyle #4}  \ppe & {\scriptstyle #5} \ppe &
 {\scriptstyle #6} \ew \end{array} \displaystyle{\} } }
\def\Mat{{\mbox{\rm Mat}}}

\def\ik{{\sf k}}
\def\min{{\mbox{\rm min\/}}}
\def\cH{{\cal H}}

\def\tr{{\rm tr}}
\def\cS{{\cal S}}
\def\tF{{\rm F}}
\def\tCS{{\rm CS}}
\def\tf{{f}}
\def\tL{{\rm L}}

\def\tA{{\rm A}} 
\def\astk{\, , \, } 

\title{{\bf Open Strings and Non-commutative} \\[3mm]  
 {\bf Geometry of Branes on Group Manifolds \footnote{Talk given by 
V.\ Schomerus at the Euroconference {\sl Brane New World and Noncommutative
Geometry}, Villa Gualino, Torino, Italy, October 2000}}\\[7mm] }
\author{  {\sc Anton Yu.\ Alekseev$\,$ \rnote{1}\ ,  
           \  Andreas Recknagel$\,$ \rnote{2} }  \\[2mm]
          {\sc   Volker Schomerus$\,$ \rnote{3}} \\[11mm]  
\note{1} Institute for Theoretical Physics, Uppsala University  
\\ Box 803, S--75108 Uppsala, Sweden
\\[2mm]
\note{2} Department of Mathematics, King's College London, \\
 Strand, London WC2R 2LS, UK \\[2mm]
\note{3} Max-Planck-Institut f\"ur Gravitationsphysik, 
       Albert-Einstein-Institut 
\\ Am M\"uhlenberg 1, D--14424 Potsdam, Germany
}
\date{April 5, 2001}
\begin{document}
\baselineskip=17pt
\begin{titlepage}      \maketitle       \thispagestyle{empty}

\vskip5mm
\begin{abstract}
In this contribution we review some recent work on the non-commutative 
geometry of branes on group manifolds. In particular, we show how fuzzy 
spaces arise in this context from an exact world-sheet description and 
we sketch the construction of a low-energy effective action for massless 
open string modes. The latter is given by a combination of a Yang-Mills 
and a Chern-Simons like functional on the fuzzy world-volume. It can be 
used to study condensation on various brane configurations in curved 
backgrounds.     
\noindent  
\end{abstract}
\vspace*{-20.9cm}
{\tt {hep-th/0104054  \hfill  AEI 2001-19}} \break
\bigskip\vfill
\noindent\phantom{wwwx}{\small e-mail: }{\small\tt alekseev@teorfys.uu.se, 
  anderl@mth.kcl.ac.uk,}\\  
  \phantom{{wwwx}{\small e-mail: }}{\small\tt vschomer@aei.mpg.de} 
\end{titlepage}

\section{Introduction} 

The connection between branes, open strings and non-commutative 
geometry has received a lot of attention recently. Almost all of 
these recent studies focused on branes in a flat background 
with constant B-field. In this case, the brane's world-volume 
geometry is given by a Moyal-Weyl deformation of the classical 
algebra of functions on the brane \cite{DouHul,ChHo,Vol} and 
scattering amplitudes of massless open string modes give rise
to a non-commutative Yang-Mills theory \cite{SeiWit}. 
\smallskip

It is of obvious interest to generalize these findings to branes 
in non-trivial backgrounds. But even though the perturbative 
analysis of \cite{Vol} suggests a very general relation between 
brane geometry and quantization theory, not much progress has been 
made in terms of background independent investigations. At the moment, 
new insights can be expected only from those backgrounds that admit an 
exact world-sheet description. For branes in flat space the latter 
involves 2-dimensional free field theories. Non-trivial backgrounds, 
however, require the use of conformal field theory (CFT) techniques. 
\smallskip

In this paper, we 
 review results from a series of papers on the 
geometry of branes on group manifolds $G$ \cite{AlSc,AlReSc1,
AlReSc2}. Our discussion will deal primarily with $G={\rm SU(2)}$, 
 but all statements have an obvious generalization to other 
groups. While $G={\rm SU(2)} \cong S^3$ appears as part of the 
NS5-brane geometry and in $AdS_3 \times S^3 \times M_4$, 
the interest in other group manifolds is more indirect 
and rests mainly on CFT model building which has its roots 
in the WZW model.  
\smallskip

There is one aspect of branes on group manifolds that deserves 
to be stressed here. Namely, group manifolds are curved and 
carry a non-vanishing NSNS 3-form field $H = dB$. This does 
not only imply that all branes carry non-vanishing B-fields 
(so that one expects to find non-commutative geometries) but
it also represents an interesting deformation of brane geometry 
in flat space. As we shall show below, the presence of a NS-field 
$H$ has intriguing consequences on brane dynamics which are similar 
to the phenomena found in \cite{Mye}. The main advantage of our 
scenario below is that it admits a full perturbative string theory 
description so that stringy effects can be taken into account.   
\smallskip

This short note is organized as follows: In the next section 
we collect some basic mathematical background on the fuzzy 
(non-commutative) geometry of certain group orbits. Based on a
semi-classical analysis, we shall then argue in Section 3 that 
these fuzzy orbits arise from branes on group manifolds in a particular 
decoupling limit. Even though this discussion makes (illegitimate)
use of some formulas that were established for branes in flat 
space, the results can be confirmed by an exact conformal field
theory construction. We review this in Section 4 before sketching 
the computation of gauge theories on the branes' world-volumes
in the final section.

\section{Orbits, quantization, and fuzzy geometry}
\def\rmc{{\rm c}}
\def\Lieg{{\cal G}}
\def\Fun{{\rm Fun}}
\def\rmY{{\rm Y}}
\def\rmC{{\rm C}} 

Our discussion of branes on group manifolds below 
will lead us to some famous quantization problem that 
has been analyzed extensively. It is useful to 
recall some of this mathematical background before
showing  
how it arises in a string theoretic context.  
To make our discussion as explicit as possible, we 
shall restrict our presentation to the example
of ${\rm SU(2)}$. The generalization to other compact 
groups is mostly obvious (but see \cite{Hop} for 
details).
\medskip 

Consider the 3-dimensional Lie algebra $\Lieg = {\rm su(2)}$.
Its structure constants $f_{ab}^{\ \ c}$ define a linear 
Poisson structure on the space $\R^3$. In terms of the 
coordinate functions $y_a$ on $\R^3$ it is given by  
\be \{ \, y_a\, ,\, y_b \, \} \ = 
                      \ f_{ab}^{\ \ c} \ y_c\ \ . 
\label{LPB1} \ee
This Poisson algebra has a large center. In fact, any 
function of $c(y) = \sum_a y_a^2$ has vanishing Poisson 
bracket with any other function on $\R^3$ so that 
formula (\ref{LPB1}) induces a Poisson structure on 
the 2-spheres 
\be c(y) \ := \ \sum_a \, y_a^2 \ \stackrel{!}{=} \ \rmc 
\label{LPB2} \ee 
of points which have the same distance $\rmc$ from the 
origin in $\R^3$. These 2-spheres are the Poisson spaces 
we want to consider. In the case $\rmc =0$ the 2-sphere
degenerates to a single point.  
\medskip

Quantization of this structure requires to find some operators 
$\rmY_a$ acting on a state space $V$ such that \vspace*{-1mm} 
\ba  & & [ \, \rmY_a \, ,\, \rmY_b \, ]\ =  
          \ i\, f_{ab}^{\ \ c} \ \rmY_c 
\label{QLPB1} \\[2mm] 
  & & \rmC \ := \ \sum\   \rmY_a^2  \ = \ 
 \ \rmc\  {\bf 1}  
\label{QLPB2}\ea   
where ${\bf 1} $ denotes the identity operator on the state
space $V$. These two requirements are the quantum 
analogues of the classical relations (\ref{LPB1}, 
\ref{LPB2}). The quantization problem posed in rel.\
(\ref{QLPB1},\ref{QLPB2}) is easy to solve. By the 
commutation relation (\ref{QLPB1}), the operators $\rmY_a$ 
have to form a representation of ${\rm su(2)}$. Condition (\ref{QLPB2}) 
states that in this representation, the quadratic Casimir 
element $\rmC$ must be proportional to the identity ${\bf 1}$. 
This is true if and only if the representation is 
irreducible. Hence, the irreducible representations
of ${\rm su(2)}$ provide quantizations of our 2-spheres.        
\smallskip

Irreducible representations of ${\rm su(2)}$ are labeled
by one discrete parameter $\a = 0, 1/2, 1,$ $\dots $. 
This implies that only a discrete set of 2-spheres 
in $\R^3$ can be quantized with their radii being 
related to the value of the quadratic Casimir in 
the corresponding irreducible representation. For 
each quantizable 2-sphere $S^2_\a \subset \R^3$ we 
obtain a state space $V^\a$ of dimension $\dim V^\a 
= 2\a + 1$ equipped with  an action of the quantized 
coordinate functions $\rmY_a$ on $V^\a$. The latter 
generate the matrix algebra $\Mat (2\a +1)$.  
\medskip

It is useful to go a bit deeper into exploring these 
quantized 2-spheres. Let us start by recalling that 
the space $\Fun(S^2)$ of functions on a 2-sphere is 
spanned by spherical harmonics $Y^J_m \in \Fun (S^2)$ 
where $J$ runs through all integer isospins. A product 
of any two spherical harmonics is again a function on 
the 2-sphere and hence it can be written as a linear 
combination of spherical harmonics, 
\be  Y^I_l \ Y^J_m \ = \ \sum_{K,n} \ 
c_{IJK}\  \CG{I}{J}{K}{l}{m}{n} \ Y^K_n 
\label{SHM} \ee 
with $[:::]$ denoting the Clebsch-Gordan coefficients 
of ${\rm su(2)}$. The structure constants $c_{IJK}$ can 
be found e.g.\ in \cite{Hop}. 
\smallskip

The algebra $\Fun(S^2)$ admits an action of ${\rm su(2)}$ which 
is obtained from the adjoint action $t_a: y_b \mapsto
f_{ab}^{\ \ c} y_c $ of ${\rm su(2)}$ on $\R^3$. It is easy to 
see that this action preserves the constraint $c(y) = \rmc$ 
and hence it descends to $\Fun(S^2)$. Since the  spherical  
harmonics $Y^1_a \in \Fun(S^2)$ are obtained by restricting
the coordinate functions $y_a$ from $\R^3$ to $S^2 \in \R^3$, 
they transform as $Y^1_b \mapsto f_{ab}^{\ \ c}Y^1_c$ under 
the action of the generator $t_a \in {\rm su(2)}$. Spherical 
harmonics $Y^J_m$ form multiplets with respect to the ${\rm su(2)}$ 
action on $\Fun(S^2)$. This classical action survives the 
quantization, i.e.\ there exists an analogous action of 
${\rm su(2)}$ on $\Mat(2\a +1)$. It is defined by 
$$ t_a: \ {\rm A} \ \mapsto\  [\, t_a^\a\, , \, {\rm A}\, ] \ \ \ 
\mbox{ for all } \ \ \ {\rm A} \in \Mat(2\a+1)\ , $$
where $t^\a_a$ are the generators of ${\rm su(2)}$ evaluated in 
the $(2\a+1)$-dimensional irreducible representation. In 
particular, it follows from the identification $\rmY^1_b := 
\rmY_b \sim t^\a_b$ that $t_a:\rmY^1_b \mapsto i f_{ab}^{\ \ c} 
\, \rmY_c^1$ as in the classical case.  
\smallskip

Under the ${\rm su(2)}$-action we have just described, the space 
$\Mat (2\a+1)$ decomposes into multiplets which are spanned by 
matrices $\rmY^J_m$ where $J = 0,1, \dots, 2\a$. The product of 
any two such matrices can be expressed as a linear combination 
of matrices $\rmY^K_n$, 
\be \rmY^I_l \ \rmY^J_m \ = \ \sum_{K \leq 2\a,n}\ 
\SJS{I}{J}{K}{\a}{\a}{\a}\, \CG{I}{J}{K}{l}{m}{n}\ 
 \rmY^K_n \ \ . 
\label{QSHM} \ee 
Here $\{:::\}$ denote the recoupling coefficients (or 6J-symbols) of 
${\rm su(2)}$. This relation can be considered as a quantization of the 
expansions (\ref{SHM}) and the classical expression is recovered 
from eq.\ (\ref{QSHM}) upon taking the limit $\a \rightarrow \infty$. 
Hence, the matrices $\rmY^J_m$ in the quantized theories  
are a proper replacement for spherical harmonics. Note, however, 
that the angular momentum $J\leq 2\a$ is bounded from above. This 
may be interpreted as `fuzzyness' of the quantized 2-spheres on 
which short distances cannot be resolved \cite{Mad}. We shall 
eventually refer to  $\rmY^J_m$ as `fuzzy spherical harmonics'. 

\section{Branes on Lie groups -- semi-classical geometry} 
\def\rmA{{\rm A}}

Strings moving on a 3-sphere $S^3$ of radius $R \sim 
\sqrt{\ik}$ are described by the SU(2) WZW model at level 
$\ik$. By the string equations of motion the 3-sphere
comes equipped with a constant NS 3-form field 
strength $ H \sim 1/\sqrt{\ik}\  \Omega$ where $\Omega$ 
denotes the volume form of the unit sphere.
\smallskip

The world-sheet swept out by an open string in $S^3$ is 
parametrized by a map $g: {\rm H} \rightarrow {\rm SU(2)}$ 
from the upper half-plane H into the group manifold SU(2)$\, 
\cong S^3$.  We shall be interested in maximally symmetric 
D-branes on SU(2) that are characterized by imposing the 
condition $J(z ) = \bar J (\bar z)$ on chiral currents all
along the boundary $z = \bar z$, i.e.
\be  - \ik \, (\partial g) g^{-1} \ = :\ J(z)  \ \stackrel{!}{=}
\Big|_{\ _{\!\!\!{\scriptscriptstyle z =\bar z}}} \  
 \bar J(\bar z) \ := \ \ik \, g^{-1} \bar\partial g\ \ . 
\label{GC}  \ee
As we shall discuss below, there exists an exact 
solution of the boundary WZW model with rel.\ (\ref{GC}). Even 
though the latter essentially goes back to Cardy \cite{Car}, 
the geometrical interpretation of the condition (\ref{GC}) 
was only found in \cite{AlSc} (see also \cite{Gaw},\cite{Sta}).%
\medskip

To describe the findings of \cite{AlSc}, we split $\partial, 
\bar \partial$ into derivatives $\partial_x, \partial_y$ 
tangential and normal to the boundary and rewrite eq.\ 
(\ref{GC}) in the form 
\be \left( \Ad(g)-1 \right)  g^{-1} \partial_y g \ = 
\Big|_{\ _{\!\!\!{\scriptscriptstyle z =\bar z}}} \  
     \, i \,\left( \Ad(g)+1 \right)  g^{-1} \partial_x g \ \ . 
\label{GC2} \ee 
Here, $\Ad(g)$ denotes the adjoint action (i.e.\ action by 
conjugation) of SU(2) on the Lie algebra {\rm su(2)}. Let us now 
decompose the  tangent space $T_h{\rm SU}(2)$ at each point $h 
\in \,$SU(2) into a part $T^{||}_h{\rm SU}(2)$ tangential to 
the conjugacy class through $h$ and its orthogonal complement 
$T^\perp_h {\rm SU}(2)$ (with respect to the Killing form). 
Using the simple fact that $\Ad_g|_{T^\perp_g} = 1$ we can 
now see that with condition (\ref{GC})  
\begin{enumerate} 
\item  
 the endpoints of open strings on SU(2) are forced to move 
 along conjugacy classes, i.e.\ 
  $$ (g^{-1} \partial_x g)^\perp \ = \ 0 \ \ .$$
 Except for two degenerate cases, namely the points $e$ and 
 $-e$ on the group manifold, the conjugacy classes are 
 2-spheres in SU(2). 
\item the branes wrapping conjugacy classes of SU(2) carry 
 a B-field which is given by 
 \be B \ \sim  \frac{\Ad(g)+1}{\Ad(g)-1} \ \ . 
 \label{Bfield} \ee 
 The associated 2-form is obtained as $\tr (g^{-1} dg \, B 
 \, g^{-1} dg)$  
 and it is easily seen to provide a potential for the NSNS
 3-form $H$.
\end{enumerate} 
The second statement follows from eq.\ (\ref{GC2}) by comparison 
with the usual condition $g_{\nu \mu}\,  \partial_y X^\mu = i\,  
B_{\nu \mu}\,  \partial_x X^\mu$. It was shown in \cite{BaDoSc},
\cite{Paw} that the spherical branes are stabilized by the 
NSNS background field $H$.      
\medskip

We know from the theory of branes in flat space that the relevant 
object that determines the geometry of branes is not $B$ but 
another anti-symmetric object $\Theta$ \cite{DouHul,ChHo,Vol}. 
Employing the standard relation between $\Theta$ and $B$ from 
flat space, we obtain 
$$
\Theta \ =\ \frac{2}{B-B^{-1}}\ =\ 
\frac{1}{2}\; \bigl(\,\Ad(g^{-1}) - \Ad(g)\,\bigr)\ .
$$
Let us have a closer look at $\Theta$ in the limit $\ik \to \infty$ 
where the 3-sphere grows and approaches flat 3-space $\R^3$. One 
can parametrize points on SU(2) by an object $X$ taking values 
in the Lie algebra su(2), such that $g \approx 1 + X$. Insertion 
into our formula for $\Theta$ gives
\be \Theta \ = \ - \ad(X) \label{ThX} \ee
where $\ad$ denotes the adjoint action of su(2) on itself. 
If we expand $X = y_c t^c$ we can evaluate the matrix 
elements of $\Theta$ more explicitly,  
$$ \Theta_{ab} \ = \  - \ (\, t_a\, ,\,  \ad(X) t_b\, ) 
   \ = \ -\, y_c\, (\, t_a\, , \, f_{\ b}^{c \ d}\, t_d\, )
   \ = \ f_{ab}^{\ \ c}\, y_c \ \ , $$       
with $(\cdot,\cdot)$ being the Killing form on su(2). Hence, the 
Poisson structure defined by $\Theta$ is determined by the 
linear Poisson bracket (\ref{LPB1}) that we discussed in the
previous section.  
\smallskip

Recall that the Moyal-Weyl products that show up for brane geometry 
in flat space with constant B-field are obtained from the constant 
Poisson bracket $\{x_\mu, x_\nu\} = \Theta_{\mu\nu}$ on $\R^d$ 
through quantization. When combined with our semi-classical analysis 
for branes on SU(2) $\cong S^3$ this suggests that the quantization 
of 2-spheres in $\R^3$ with Poisson bracket (\ref{LPB1}) becomes 
relevant for the geometry of branes on ${\rm SU(2)}$ in the limit 
where $\ik \rightarrow \infty$. Consequently, we expect the 
world-volume of branes on SU(2) to be  fuzzy 2-spheres. This will 
be confirmed by the exact construction of open string backgrounds 
in the next section.

\section{Branes in WZW models -- the exact solution}

As explained in \cite{Vol,AlReSc1}, the geometry of branes can be 
read off from the operator product of tachyonic open string
vertex operators; see also \cite{FrGaw} for earlier such proposals 
in the closed string context. This procedure requires to solve the 
underlying world-sheet theory. In the theory of branes in 
flat space, Wick's theorem applies so that  all information 
about the boundary theory is encoded in the propagator. 
This is no longer true for non-trivial backgrounds such 
as group manifolds. To deal with such more general 
situations, one employs techniques of boundary conformal 
field theory. We will give a short survey of the 
constructions relevant for WZW models. 
\medskip
     
The chiral fields of the SU(2) WZW model form an affine 
Kac-Moody algebra denoted by $\widehat{\rm SU}(2)_\ik$. It 
is generated by fields $J_a(z),\ a = 1,2,3,$ which possess 
the following operator product expansions
\be  
J_a (x_1) \ J_b(x_2)\  = \ \frac{\ik}{2}\, \frac{ \delta_{a b}}
                            {(x_1-x_2)^2}
         \ + \  \frac{i\, f_{ab}^{\ \ c}}{(x_1-x_2)} \, J_c(x_2) 
           \ + \ \dots \ \ . \ \label{JOPE1} 
\ee
The current $J(z)$ from Section 3 is given by $J = J_a t^a$. 
For the situation we are dealing with (gluing conditions $J = 
\bar J$ in a ``parent'' CFT on the full complex plane with 
diagonal modular invariant partition function), Cardy \cite{Car} 
was able to list all possible boundary conditions. There exist 
$\ik +1$ of them, differing in the bulk field one-point functions 
(brane charges) and labeled by an index $\a = 0, \frac{1}{2}, 
\dots, \frac{\ik}{2}$. Without entering a detailed description 
of these boundary theories \cite{Car}, we recall that their state 
spaces have the form 
\def\cH{{\cal H}}
\be \label{partdec}
 \cH_\a \ = \ {\bigoplus}_J \ N_{\a\a}^J \ \cH^J 
\ee
where $\cH^J$, $J= 0,1/2, \dots,\ik/2$, denote 
irreducible highest weight representations of the affine Lie algebra 
${\widehat{{\rm SU}}(2)}_\ik$, and where $N_{IJ}^K$ are the associated 
fusion rules. Note that only integer spins $J$ appear on the right hand 
side of (\ref{partdec}). 

\smallskip
There exists a variant of the state-field correspondence which 
assigns a boundary field $V(x)$ to each element $|V\rangle \in 
\cH_\a$ (see e.g.\ \cite{ReSh1}). Each representation $J$ 
contains ground states of lowest possible energy, which form 
$(2J+1)$-dimensional multiplets spanned by some basis vectors 
$\rmY^J_m$ with $|m|< J$. To these states we assign vertex operators 
$V[\rmY^J_m](x)$. They are similar to the tachyonic vertex operators 
$V[e_k] = \exp (-ik X)$ for open strings in flat space. While the 
latter are associated with eigenfunctions of linear momentum $k$, 
our vertex operators $V[\rmY^J_m]$ carry definite `angular' 
momentum, i.e.\ the obey 
\be\label{Ldef}
J_a (x_1) \ V[\rmY^J_m](x_2) \ = \  
\frac{1}{x_1-x_2} \ V[ L_a \rmY^J_m] (x_2) \ + \ \dots  
\ee
where $L_a \rmY^J_m = (t^J_a)_{mn} \rmY^J_n$ denotes the action 
of su(2) on the multiplet $(\rmY^J_j)$. Eq.\ (\ref{partdec}) 
shows that the angular momentum is cut off at $J = \min(2\a, 
\ik- 2\a)$. In the limit $\ik \rightarrow \infty$, this cut-off 
agrees with the one that we encountered in our discussion of 
fuzzy 2-spheres. In other words, for $\ik \rightarrow \infty$ 
the tachyonic open string vertex operators in the boundary 
theory labeled by $\a$ are in one-to-one 
correspondence with 
fuzzy spherical harmonics of the fuzzy 2-sphere $S^2_\a$. This
can be regarded as the first confirmation of the expectation 
formulated at the end Section 3. 
\medskip

The operator product expansion of these open string vertex 
operators are subject to certain factorization constraints
(associativity). They were formulated first by 
Lewellen \cite{Lew} and solved for minimal models by 
Runkel in \cite{Run}. In case of the WZW boundary theories, 
the solution reads \cite{Run,AlReSc1}  
\be \label{boundOPE}
    V[\rmY^I_m](x_1)\, V[\rmY^J_n](x_2) \, = \, {\sum}_{K,k}\, 
     x_{12}^{h_K - h_I -
    h_J} \, \CG{I}{J}{K}{i}{j}{k}\, \Fus{\alpha}{K}{J}{I}{\alpha}
    {\alpha} \, V[\rmY^K_k](x_2) + \dots     
\ee
where $h_J = J(J+1)/(\ik+2)$ is the conformal dimension of 
$V[\rmY^J]$ and $F$ stands for the fusing matrix of the WZW model. 
In the limit $\ik \rightarrow \infty$, the fusing matrix elements 
approach the $6J$ symbols of the classical Lie algebra ${\rm su}(2)$. 
At the same time the conformal dimensions $h_J = J(J+1)/(\ik+2)$ 
tend to zero so that the OPE (\ref{boundOPE}) of boundary fields 
becomes regular as in a topological model. In this sense, the 
limit $\ik \rightarrow \infty$ is similar to the decoupling 
limit considered in \cite{SeiWit}.  
 
Using the last two observations about the limiting theory we
obtain a much more elegant way of writing the operator product 
expansion (\ref{boundOPE}). In fact, the comparison of eq.\ 
(\ref{boundOPE}) with eq.\ (\ref{QSHM}) shows that the operator 
product expansion is encoded in the multiplication of matrices  
if we think of $\rmY^J_m$ as elements of $\Mat(2\a + 1)$, i.e.\ 
for $\ik \rightarrow \infty$ we find  
$$     
    V[\rmY^I_m](x_1)\ V[\rmY^J_n](x_2) \ = \ 
    V[\rmY^I_m  \, \rmY^J_n](x_2)
    \ \ \ \mbox{ for } \ \ \rmY^I_m, \rmY^J_n  
    \in \Mat(2\a+1)\ \  
$$
up to subleading terms. 
The emergence of the matrix product in this relation confirms 
the results from Section 3 since it shows that the world-volumes 
of branes in $S^3$ become fuzzy two-spheres when we send the 
level to infinity. We can now rewrite the operator product in 
a way which does no longer refer to a particular choice of  
basis in $\Mat(2\a+1)$. For an arbitrary matrix $\rmA \in 
\Mat(2\a+1)$ with $\rmA = \sum a_{J m} \rmY^J_m$ we introduce 
$V[\rmA] = \sum a_{Jm} V[\rmY^J_m]$ and obtain 
\be \label{boundOPE2}
 V[\rmA_1](x_1)\ V[\rmA_2](x_2) \ = \ V[\rmA_1 \, \rmA_2](x_2)\ \ \   
\ee
for all $\rmA_1,\rmA_2 \in \Mat(2\a+1)$. This product allows us to 
compute arbitrary correlations functions of such vertex operators, 
\be \langle V[\rmA_1](x_1) \ V[\rmA_2](x_2) \ \cdots \ 
  V[\rmA_n](x_n) \rangle \ = \ \tr (\rmA_1 \ \rmA_2 \  \cdots \  \rmA_n)\ \ .
\label{wzwcorr} \ee
The trace appears because the vacuum expectation value is  SU(2) 
invariant and the trace maps matrices to their SU(2) invariant 
component. Formulas (\ref{JOPE1},\ref{Ldef},\ref{wzwcorr}) provide 
a complete solution of the boundary WZW models in the decoupling 
limit $\ik \rightarrow \infty$. The generalization to branes 
localized along conjugacy classes of other Lie groups is 
straightforward (see \cite{FFFS} for details).  

\section{Open strings and gauge theory on fuzzy orbits}
 
Equipped with  
the exact solution of the boundary WZW model we are 
finally prepared to calculate the low-energy 
effective action 
for massless open string modes. Compared to the flat space 
case \cite{SeiWit} there are two important changes in the 
computation. First, it follows from eq.\ (\ref{wzwcorr}) that 
Moyal-Weyl products get replaced by matrix multiplication. 
Second, there appears a new term $f_{ab}^{\ \ c} J_c$ in 
the operator product expansion of currents (\ref{JOPE1}). 
This term leads to an extra contribution of the form 
$f_{abc} \rmA^a \rmA^b \rmA^c$ in the scattering amplitude 
of three massless open string modes. Consequently, the 
resulting effective action is not given by Yang-Mills 
theory on a fuzzy 2-sphere but  involves also a 
Chern-Simons like term. 
\smallskip

For $M$ branes of type $\a$ on top of each other, the results 
of the complete computation \cite{AlReSc2} 
can be summarized in the following 
formula, 
\be  \label{effact}
  \cS_{(M , \a)}\ =\  \cS_{{\rm YM}} + \cS_{{\rm CS}}\ = \
      \frac{1}{4}\ \tr \left( \tF_{a b} \ \tF^{a b} \right) 
             - \frac{i}{2}\  \tr \left( \tf^{a b c}\; \tCS_{a b c} \right)
\ee
where we defined the `curvature form' $\tF_{ab}$ by the expression     
\be \label{fieldstr}
  \tF_{a b}(\tA)\, = \,   
   i\, \tL_a \tA_b - i\, \tL_b \tA_a + i \,[ \tA_a \astk \tA_b] 
   +  \tf_{a b c } \tA^c
\ee
and a non-commutative analogue of the Chern-Simons form by 
\be  \label{CSform}
  \tCS_{a b c}(\tA)\, = \, \tL_a \tA_b \, \tA_c 
                   + \frac{1}{3}\; \tA_a \, [ \tA_b \astk \tA_c]
                   - \frac{i}{2}\; \tf_{a b d}\; \tA^d \, \tA_c \ \ .
\ee
The three fields $\tA^a = \sum a^a_{Jm} \rmY^J_m$ on the fuzzy 
2-sphere $S^2_\a$  take values in $\Mat(M)$, i.e.\ $a^a_{Jm}
\in \Mat(M)$. Hence the fields $\tA^a$ are elements of
$\Mat(M) 
\otimes \Mat(2\a+1)$. Gauge invariance of (\ref{effact}) under 
the gauge transformations 
$$ \tA_a  \ \rightarrow \ L_a \Lambda \ +\  i\, [\, A_a\, ,\, 
    \Lambda\, ]  \ \ \ \mbox{ for } \ \ \ \Lambda \in 
    \Mat(M) \otimes \Mat(2\a+1)  
$$
follows by straightforward computation. Note that the 'mass term' 
in the Chern-Simons form (\ref{CSform}) guarantees the gauge 
invariance of $\cS_{{\rm CS}}$. On the other hand, the effective 
action (\ref{effact}) is the unique combination of $\cS_{\rm YM}$ 
and $\cS_{\rm CS}$ in which mass terms cancel. As we shall see below, 
it is this special feature of our action that allows solutions 
describing translations of the branes on the group manifold. The 
action $\cS_{\rm YM}$ was already considered in the non-commutative 
geometry literature \cite{FFT,Madbook,Wat}, where it was derived from 
a Connes spectral triple and viewed as describing Maxwell theory on 
the fuzzy sphere. Arbitrary linear combinations of non-commutative 
Yang-Mills and Chern-Simons terms were considered in \cite{Klim}.  
\medskip

{}From eq.\ (\ref{effact}) we obtain the following equations of 
motion for the elements $\tA^a \in \Mat(M) \otimes \Mat(2\a+1)$   
\begin{equation} 
\label{eom} 
\Bigl[\, A^a  \;\, , [\, A_a \, , \,  A_b\, ] \, - \,  
 i\, f_{a b c}\, A^c \,\Bigr]  \ = \ 0 \ \ . 
\end{equation} 
Solutions of these equations (\ref{eom}) describe condensates 
on a stack of $M$ branes of type $\a$.  It is easy to find 
two very different types of solutions. The first one is given by 
a set of $3$ pairwise commuting $M \times M$ matrices $A_a$. 
It comes as an $M \cdot 3 $ parameter family of solutions 
corresponding to the number of eigenvalues appearing in $\{ A_a\}$. 
The same kind of solutions appears also for branes in flat backgrounds. 
They describe  rigid translations of the $M$ branes on the 
group manifold. Since each brane's position is specified by $3$ 
coordinates, the number of parameters matches nicely with the 
interpretation. Moving branes around in the background is a rather 
trivial symmetry and the corresponding conformal field theories 
are easy to construct (see also \cite{ReSc2}). Note, however,  
that the existence of these solutions is guaranteed by the 
absence of the mass term in the full effective action.  
\smallskip

There exists a second type of solutions to eqs.\ (\ref{eom}) 
which is a lot more interesting. In fact, any $M(2\a + 1)$-dimensional
representation of the Lie algebra su(2) can be used to solve
the equations of motion. Their interpretation was found in 
\cite{AlReSc2}. Let us describe the answer for a stack of 
$M$ branes of type $\a = 0$, i.e.\ of $M$ point-like branes 
at the origin of SU(2). In this case, $\tA_a \in \Mat(M) 
\otimes \Mat(1) \cong \Mat(M)$ so that we need an $M$-%
dimensional representation of su(2) to solve the equations
of motion. Let us choose the $M$-dimensional irreducible 
representation $\sigma$. Our claim then is that this drives 
the initial stack of $M$ point-like branes at the origin into 
a final configuration containing only a single brane 
wrapping the sphere of type $\a = (M-1)/2$, i.e.\ 
$$ (M,\,\a = 0 )  \ \stackrel{\sigma}{\longrightarrow} \ 
   (1,\,\a = (M-1)/2) \ \ . $$
Support for this statement comes from both the open string 
sector and the coupling to closed strings (see \cite{AlReSc2}). 
In the open string sector one can study small fluctuations 
$\delta A_a$ of the fields $A_a = \Lambda_a + \delta A_a  
\in \Mat(M)$ around the stationary point $\Lambda_a
\in \Mat(M)$. If $\Lambda_a$ form an irreducible representation
of su(2), one finds, 
$$  \cS_{(M,0)} (\Lambda_a + \delta \tA_a) \ = \ 
    \cS_{(1,(M-1)/2)}(\delta \tA_a) \ + \ \mbox{const} \ \ \ .  
$$ 
In the closed string channel one can show, along the lines of 
\cite{ReRoSc}, that the leading 
term (in the 1/\ik -expansion) from the exact mass formula 
of a brane wrapping the sphere of type $\a = (M-1)/2$ 
coincides with the value of the action $\cS_{(M,0)}
(\Lambda_a)$ at the stationary point $\Lambda_a$ 
\cite{AlReSc2}. Note that the mass of the final state
is lower than the mass of the initial configuration. This 
means that a stack of $M$ point-like branes on a 3-sphere 
is unstable against decay into a single spherical brane. 
Stationary points of the action (\ref{effact}) and the 
formation of spherical branes on $S^3$ were also discussed 
recently in \cite{HasKra,HiNoSu}. Similar effects 
have been described for branes in RR-background fields 
\cite{Mye}. The advantage of our scenario with NSNS-background
fields is that it can be treated in perturbative string
theory so that string effects may be taken into account
(see \cite{FreSch}).   

\section{Conclusion and Outlook}

The analysis we have reviewed here generalizes the well-known 
relation between open strings and non-commutative
geometry of branes in flat space to a large class of 
branes on group manifolds. In particular, we have 
described how fuzzy geometries and gauge theories 
on these spaces  
appear in this context. Our main result 
(\ref{effact},\ref{fieldstr},\ref{CSform}) was formulated 
for su(2) but its generalization to other Lie algebras 
is essentially obvious. 
\smallskip

The condensation processes that are encoded in the effective
action were used in \cite{AlSc2} to argue that the charges
of branes on $S^3$ are only defined modulo some integer. This 
seems to fit nicely \cite{FreSch} with a proposal of Bouwknegt 
and Mathai in \cite{BouMat} according to which the charges of branes 
in a background $X$ with non-vanishing NSNS H-field $H \in 
H^3(X,\Z)$  generate certain twisted K-theory groups $K_H(X)$. 
\smallskip

Let us finally recall that WZW models provide the most 
important starting point for CFT model building, such 
as coset and orbifold constructions. Hence, one can hope
to use the results reported above to analyze brane dynamics
in many other backgrounds, such as minimal models, various
Kazama-Suzuki quotients and Gepner models. The concrete 
realization of this program for $N=2$ supersymmetric 
minimal models is sketched in \cite{FreSch2}.  
\bigskip
\bigskip

\def\gaw{Gawedzki}
\newcommand{\sbibitem}[1]{\vspace*{-1ex} \bibitem{#1}}

\end{document}